\documentclass[runningheads]{llncs}

\usepackage[hidelinks]{hyperref} 
\usepackage{graphicx} 
\usepackage{amsfonts} 
\usepackage[bottom]{footmisc} 
\usepackage{subfig}
\usepackage{verbatim} 
\usepackage{enumitem} 
\usepackage{algorithm}
\usepackage{todonotes}
\usepackage[noend]{algpseudocode}
\usepackage{cite}

\makeatletter
\def\BState{\State\hskip-\ALG@thistlm}
\makeatother

\begin{document}
\pagestyle{headings}

\title{Reactive Proximity Data Structures for Graphs}
\author{David Eppstein \and Michael T. Goodrich \and
Nil Mamano}
\authorrunning{D. Eppstein \and M.T. Goodrich \and N. Mamano}
\institute{Department of Computer Science, University of California, Irvine, USA\\
\email{\{eppstein,goodrich,nmamano\}@uci.edu}}

\maketitle             

\begin{abstract}
We consider data structures for graphs where we maintain
a subset of the nodes called \emph{sites}, and allow proximity 
queries, such as asking
for the closest site to a query node, and update operations that
\emph{enable} or \emph{disable} nodes as sites. 
We refer to a data structure that can efficiently react to such updates as
\emph{reactive}. We present novel reactive 
proximity data structures for graphs of polynomial expansion, i.e., the class of graphs with small separators, such as planar
graphs and road networks.
Our
data structures can be used directly in several logistical problems and geographic information systems dealing with real-time data, such as emergency dispatching.
We experimentally compare our data structure to Dijkstra's algorithm in a system emulating random queries in a real road network.
\end{abstract}

\section{Introduction}

\textit{Proximity data structures} maintain a set of objects of interest, called \textit{sites}, and support queries concerned with minimizing some distance involving the sites, such as nearest-neighbor or closest-pair queries.
They are well known in computational geometry~\cite{de2008computational}, where sites are points in space and distance is measured by Euclidean distance or some other metric (e.g., see~\cite{knuth1998art,AgaEppMat-FOCS-92,Chan2010,KapMulRod-16,chan:LIPIcs:2019:10428}).
In this chapter, we are interested in proximity data structures that deal with nodes in a graph rather than points in space.
We consider that there is an underlying, fixed graph $G$, such as a road network for a geographic region, and sites are a distinguished subset $P$ of the vertices of $G$. Distance is measured by shortest-path distance in $G$.
We consider updates (additions and deletions) to and from the set $P$ of sites. Our goal is to design efficient data structures for the following problems in graphs.

\begin{definition}[Reactive nearest-neighbor data structure]\label{def:gnndsagain}
	Given a fixed, undirected graph $G=(V,E)$ with positively-weighted edges,
	maintain a subset of nodes $P\subseteq V$, subject to insertions to $P$, deletions from $P$, and nearest-neighbor queries: given a query node $q\in V$, return the node $p\in P$ closest to $q$ in shortest-path distance.
\end{definition}

\begin{definition}[Reactive closest-pair data structure]\label{def:gcpds}
	Given a fixed, undirected graph $G=(V,E)$ with positively-weighted edges,
	maintain a subset of nodes $P\subseteq V$, subject to insertions to $P$, deletions from $P$, and queries asking for the closest pair in $P$.
\end{definition}

\begin{definition}[Reactive bichromatic closest-pair data structure]\label{def:gbcpds}
	Given a fixed, undirected graph $G=(V,E)$ with positively-weighted edges,
	maintain two subsets of nodes $P,Q\subseteq V$, subject to insertions to $P$ or $Q$, deletions from $P$ or $Q$, and queries asking for the closest pair of nodes in different sets (one in $P$ and one in $Q$).
\end{definition}

\subsection{Background}\label{gnn:sec:bg}

The data structures that we study fall into the area of \emph{dynamic graph algorithms}, the subject of extensive study~\cite{EppGalIta-ATCH-99}.
Traditionally, dynamic data structures in graphs, e.g., for shortest-path computations, allow updates on the underlying graph $G$ itself, such as vertex or edge insertions and deletions~\cite{EppGalIta-ATCH-99,Kin-FOCS-99,DemIta-JACM-04,RodZwi-Algo-11,DjiPanZar-STACS-95,ChaYan-IEEETC-09,BurResTho-IJC-08}.
We call our data structures \emph{reactive} to distinguish the kind of updates we allow. For us, $G$ is fixed, but we allow updates on $P$.

Previous work on dynamic graph algorithms has focused on the setting where $G$ can change.
Exceptions are the work of Eppstein on maintaining a dynamic subset of vertices in a sparse graph and keeping track of whether it is a dominating set~\cite{Epp-TALG-09}, and the work of Italiano and Frigioni on dynamic connectivity for subsets of vertices in a planar graph~\cite{FriIta-Algo-00}.
Despite the applications that we mention in Section~\ref{gnn:sec:apps}, to our knowledge, no one has considered 
proximity data structures for graphs subject to updates on the set of sites.

We design data structures that only work for graphs from certain hereditary graph classes.
A \textit{graph class} is a (generally infinite) set of all graphs with some defining property.
A graph class is \textit{hereditary} if it contains every induced subgraph of every graph in the class. An \textit{induced subgraph} of a graph $G$ is a graph obtained by removing any number of vertices (and their incident edges) from $G$.
For instance, the class of planar graphs is hereditary because every induced subgraph of a planar graph is planar.

Our data structures work for graphs from hereditary graph classes with separators of sublinear size.
A \textit{separator} of a graph $G=(V,E)$ is a subset of $V$ whose removal from $G$ splits $G$ into two disjoint subgraphs, each with at most $\frac{2}{3}|V|$ nodes, and with no edges between them.
We say a graph class has $O(n^c)$-size separators if every $n$-node graph in the class has a separator of size $O(n^c)$. A graph class has sublinear separators if it has $O(n^c)$-size separators for some $c<1$.
For instance, the planar separator theorem states that planar graphs have $O(n^{0.5})$-size separators~\cite{lipton1979}.

A hereditary graph class has sublinear separators if and only if it has polynomial expansion~\cite{dvorak2016}.
Thus, any graph from a class of polynomial expansion is suitable for our data structures.

If $G=(V,E)$ is a graph from a hereditary graph class with sublinear separators, then $G$ is \textit{sparse}, which means that $|E|\;=O(|V|)$. The converse is not necessarily true. For instance, bounded-degree expander graphs are sparse but do not have sublinear separators~\cite{hoory2006expander}.
Nonetheless, many important sparse graph families are hereditary and have sublinear separators.
One of the first classes that was shown to have sublinear separators is the class of planar graphs, which have $O(n^{0.5})$-size separators~\cite{lipton1979}.
Separators of the same asymptotic size have also been proven to exist for $k$-planar graphs~\cite{DujEppWoo-SJDM-17}, bounded-genus graphs~\cite{gilbert1984}, minor-closed graph families~\cite{kawarabayashi2010}, and the graphs of certain four-dimensional polyhedra~\cite{eppstein2016treetopes}.
In addition, trees have separators of size one. More generally, graphs with bounded treewidth~\cite{ROBERTSON198449} have constant-size separators~\cite{chung1989}.

The importance of having sublinear separators in our data structures is that it allows us to construct a \textit{separator hierarchy}.
A separator hierarchy is the result of recursively partitioning a graph into disjoint subgraphs using separators.
Separator hierarchies are useful to solve many graph problems~\cite{GOODRICH1995374,Frieze1992}.
An important application is the \textit{single-source shortest path} (SSSP) problem: finding the distance from a node to every other node in the graph. This problem can be solved in linear time given a type of separator hierarchy called a \textit{recursive division}~\cite{HENZINGER19973}. For graphs for which we can construct this hierarchy in linear time, such as planar graphs~\cite{HENZINGER19973}, the SSSP problem can be solved in linear time. This improves upon the $O(n\log n)$ time required by Dijkstra's algorithm in sparse graphs~\cite{Dijkstra1959}. 


In many applications (see Section~\ref{gnn:sec:apps}), the underlying graph $G$ represents a real road network. 
A road network can be represented by a graph where each node is an intersection, and each edge is a stretch of road connecting two intersections. Edge weights represent road lengths. Road networks are often modeled as planar graphs. However, they are not quite planar because of bridges and underpasses~\cite{Eppstein2008}. Thus, we are particularly interested in a class of graphs which has been shown to be a better model for road networks: the class of \textit{graphs with sparse crossing graphs}~\cite{eppstein2017crossing}.
Given an embedding of a graph $G$ in the plane, the \textit{crossing graph} of the embedding is a graph $H$ where each node in $H$ represents an edge of $G$, and two nodes in $H$ are connected if the corresponding edges in $G$ cross in the embedding. Clearly, a graph is planar if it has an embedding such that the corresponding crossing graph has no edges. More generally, it is $k$-planar if it has an embedding such that the crossing graph has maximum degree $k$. Graphs with sparse crossing graphs further generalize $k$-planar graphs: a graph has a sparse crossing graph if it has an embedding such that the corresponding crossing graph has bounded \textit{degeneracy}, a notion of sparcity used in graph theory~\cite{lick_white_1970}. Bounding the degeneracy of the crossing graph instead of the maximum degree accounts for, e.g., long tunnels that go under many street-level roads.
Like planar graphs, the class of graphs with sparse crossing graphs is also hereditary and has $O(n^{0.5})$-size separators~\cite{eppstein2017crossing}.
This is fortunate, because it means that we can use our data structures in applications dealing with road networks.

\subsection{Our contributions}
We design a new reactive nearest-neighbor data structure (Definition~\ref{def:gnndsagain}) with the aim to balance between query and update times.
If we only cared about one of these, the data structure would be trivial.
For instance, if we only cared about query time, there is a well known solution: the graph-based Voronoi diagram, which maintains the closest site to each node in the graph. Erwig~\cite{Erwig2000} shows that Voronoi diagrams can be adapted to graphs and that they can be constructed using
a modification of Dijkstra's algorithm.
With this information, queries can be answered in constant time.
However, the Voronoi diagram is not easy to update, requiring $O(n\log n)$ time in sparse graphs with $n$ nodes---the same time as for creating the diagram from scratch.

If, instead, we optimize for update time only, we could avoid maintaining any information and answer queries directly using a shortest-path algorithm from the query node.
Updates would take constant time; queries could be answered using Dijkstra's algorithm~\cite{Cormen2001}, which runs in $O(n\log n)$ time in sparse graphs. As mentioned, this could be improved to $O(n)$ time for graphs for which we can construct a recursive subdivision during a preprocessing stage~\cite{HENZINGER19973}.

Our reactive nearest-neighbor data structure finds a ``sweet spot'' between fast queries and fast updates.
Table~\ref{tab:gnnsummary} summarizes its runtime as a function of the size of the separators (the data structure is the same when $c=0$, but an extra logarithmic factor appears in the analysis). For planar graphs and, more generally, graphs with sparse crossing graphs, $c=1/2$. For graphs with bounded treewidth, $c=0$.

\begin{table}[]
	\centering
	\begin{tabular}{cccccc}
		Sep. size     & Space          & Preprocessing   & Query           & Insertion & Deletion                    \\ \hline
		$0<c<1$      & $O(n^{1+c})$   & $O(n^{1+c})$    & $O(n^c)$     & $O(n^c)$ & $O(n^c\log k)$         \\
		& $O(n^{1+c})$   & $O(n^{1+c}\log n)$  & $O(n^c)$     & $O(n^c\log\log n)$ & $O(n^c\log\log n)$     \\ \hline
		$c=0$        & $O(n\log n)$   & $O(n\log n)$    & $O(\log n)$     & $O(\log n)$  & $O(\log n\log k)$         \\
		& $O(n\log n)$   & $O(n\log^2 n)$  & $O(\log n)$     & $O(\log n\log\log n)$ & $O(\log n\log\log n)$     \\ 
	\end{tabular}
	\caption[Runtimes of the reactive nearest-neighbor data structure.]{Runtimes of our reactive nearest-neighbor data structure when it maintains $k$ sites on an $n$-node graph from a hereditary graph class with $O(n^c)$-size separators. The preprocessing time is under the assumption that a separator can be found in $O(n^{1+c})$ time. Possible trade-offs between preprocessing and update times are shown.}
	\label{tab:gnnsummary}
\end{table}

To construct a reactive nearest-neighbor data structure for planar graphs specifically, we could also consider using an \textit{exact-distance oracle}. This is a static data structure that admits queries asking for the distance between any two nodes. If $k$ is the number of sites, with an exact-distance oracle we can find the closest site to a query node with $k$ queries. The recent oracle from Gawrychowski~et al.~\cite{Gawrychowski18} answers queries in $O(\log n)$ time when it uses $O(n^{1.5})$ space like our data structure. With this oracle, we could answer queries for our data structure in $O(k\log n)$ time and do updates in constant time.
This approach has better runtimes when $k$ is small, but the preprocessing time is $O(n^2)$.

We combine our reactive nearest-neighbor data structure with preexisting data structures~\cite{Eppstein2000,Eppstein1995} to obtain other proximity data structures. Table~\ref{tab:gdssummary} shows our new family of proximity data structures. Each data structure has a similar preprocessing--update time trade-off as shown in Table~\ref{tab:gnnsummary}. For brevity, we only show the versions of the data structures that optimize the preprocessing time. 

\begin{table}[]
	\centering
	\begin{tabular}{c|ccc}
		Data structure & Query    & Insertion          & Deletion           \\ \hline
		Exact NN  & $O(n^c)$ & $O(n^c)$     & $O(n^c\log k)$     \\
		Closest pair  & $O(1)$ & $O(n^c\log^2 k)^\dagger$     & $O(n^c\log^3 k)^\dagger$     \\
		Bichromatic CP & $O(1)$ & $O(n^c\log^2 k)^\dagger$     & $O(n^c\log^3 k)^\dagger$     
	\end{tabular}
	\caption[Summary of reactive proximity data structures.]{Runtimes of our reactive proximity data structures when they maintain $k$ sites on an $n$-node graph from a hereditary graph class with $O(n^c)$-size separators, where $0<c<1$ (we omit the case of $c=0$ for brevity).
		All the data structures require $O(n^{1+c})$ space. The preprocessing time is $O(n^{1+c})$ assuming that a separator can be found in $O(n^{1+c})$ time. The ``$\dagger$'' superindex indicates that the runtime is amortized.}
	\label{tab:gdssummary}
\end{table}

\subsection{Applications}\label{gnn:sec:apps}

Reactive proximity data structures in graphs can be useful in
several logistical problems in geographic information systems dealing with real-time data.
Consider an application to connect drivers and clients in 
a private-driver service, such as Uber or Lyft, or even 
a future self-driving car service.
A reactive nearest-neighbor data structure could maintain the set of cars waiting at
various locations in a city to be put into service.
When a client requires a driver, she queries the data structure 
to find the car nearest to her. This car is then removed from $P$ (i.e., it
is no longer available) 
until it completes the trip for this client, at which point the car is then
added to $P$ (i.e., it is available) at this new location.
Alternatively,
we could consider a similar application in the context of 
police or emergency dispatching, 
where the data structure maintains the locations of a set of 
available first responder vehicles.
In Section~\ref{gnn:sec:exps}, we experiment with this type of system emulating random queries in a real road network.

In a companion paper~\cite{eppstein2017_2}, we use it for political redistricting. Suppose we are given a set of sites representing
the locations of certain facilities, such as post offices or voting locations. We
wish to partition the vertices of the graph into geographic regions, one for each
facility, such that each region has a specified size (in number of nodes) and the
shapes of the regions satisfy certain compactness criteria. As we show in the
companion paper, a greedy matching algorithm can exploit an efficient reactive
data structure to quickly build such a partitioning of the graph.
In another paper~\cite{Mamano2019ISAAC}, we use this data structure as part of an algorithm for Steiner TSP in road networks.

Reactive proximity data structures can also be useful in other domains, 
such as content distribution networks, like the one maintained by Akamai.
For instance, a reactive nearest-neighbor data structure could maintain the set of nodes that 
contain a certain file of interest, like a movie.
When another node in the network needs this information, 
the data structure could be used to find the closest node that can transfer it.
Updates allow us to model how copies of such a file
migrate in the network, e.g., for load balancing, so that we add a node
to $P$ when it gets a copy of the file and remove a node from $P$ when it passes the file to another
server.

\section{Nearest-neighbor data structure}\label{gnn:sec:algo}

Initially, we are given an $n$-node graph $G=(V,E)$ and a subset $P\subseteq V$ of sites.
As mentioned, the runtime analysis depends on the size of the separators. Henceforth, we consider that $G$ is undirected, has positive edge weights, and comes from a hereditary graph class with $O(n^c)$-size separators for some constant $c$ with $0<c<1$ (the analysis is slightly different when $c=0$).

We begin by reviewing the concept of a separator hierarchy.
Recall that a \emph{separator} in a given $n$-vertex graph is a subset $S$ of nodes such that the removal of $S$ (and its incident edges) partitions the remaining graph into two disjoint subgraphs (with no edges from one to the other), each of size at most $2n/3$.
It is allowed for these subgraphs to be disconnected; that is, removing $S$ can partition the remaining graph into more than two connected components, as long as those components can be grouped into two subgraphs that are each of size at most $2n/3$.
A \emph{separator hierarchy} is the result of recursively subdividing a graph by using separators.
Since children have size at most $2/3$ the size of the parent, the separator hierarchy is a binary tree of $O(\log n)$ height.

\subsection{Preprocessing}

The creation of our data structure consists of two phases.
The first phase does not depend on $P$, while the second phase incorporates our knowledge of $P$.
Note that there are two kinds of nodes of interest: separator nodes and sites.
The two sets may intersect, but should not be confused.

\paragraph{Site-independent phase.}
First, we build a \emph{separator hierarchy} of the graph.
This hierarchy can be constructed in $O(n)$ time and space in planar graphs~\cite{GOODRICH1995374} and graphs with sparse crossing graphs~\cite{eppstein2017crossing}. However, we do not need the construction to take linear time, as this is not the bottleneck of the preprocessing. It suffices that the hierarchy can be computed in $O(n^{1+c})$ time.
In fact, it suffices that a single separator can be found in $O(n^{1+c})$ time in an $n$-node graph (as opposed to the entire hierarchy). This is because the hierarchy is built recursively so, if a separator can be found in $O(n^{1+c})$ time, the construction time of the entire hierarchy is captured by the recurrence
\begin{equation}\label{eq:master}
T(n)\leq T(x)+T(y)+O(n^{1+c}),
\end{equation}
where $x$ and $y$ are the sizes of two subgraphs, chosen so that $x+y\le n$, $\max(x,y)\le 2n/3$, and, among all $x$ and $y$ obeying these constraints, $T(x)+T(y)$ is maximum. It is easy to prove that this recurrence is dominated by its top-level $O(n^{1+c})$ term, so $T(n)=O(n^{1+c})$.

Second, we compute, for each graph in the hierarchy, the distance from each separator node to every other node.
Consider the work done for the graph at the root of the hierarchy, $G$ itself.
We need to compute $O(n^c)$ SSSP problems, one for each separator node.
As mentioned, each such problem can be solved in linear time given a recursive subdivision~\cite{HENZINGER19973}. A recursive subdivision is a type of separator hierarchy that is also built by finding separators recursively. Thus, if we can find a separator in $O(n^{1+c})$ time, we can construct the entire recursive subdivision, and compute all the distances for the separators in the top-level graph, in $O(n^{1+c})$ time. 
We do the same for all the remaining graphs in the separator hierarchy. The total runtime follows Equation~\ref{eq:master} again, so it is also $O(n^{1+c})$.

\paragraph{Site-dependent phase.}
For each graph $H=(V_H,E_H)$ in the separator hierarchy, for each separator node $s$ in $H$, we initialize a priority queue $Q_s$.
The elements stored in $Q_s$ are the sites in $H$, $P\cap V_H$. Their priorities are their distances from~$s$ in $H$.

We use an implementation of a priority queue that supports insertions and find-minimum operations in constant worst-case time, and deletions in logarithmic worst-case time. For instance, we can use a strict Fibonacci heap~\cite{Brodal2012} or a Broadal queue~\cite{Brodal96}.
Then, constructing each queue $Q_s$ takes time linear on the number of sites in $H$.
Thus, the time at the top level of the hierarchy is $O(|P|)$ per separator node, and $|P|=O(n)$, so in total it is $O(n^{1+c})$. The total time analysis of this phase is $O(n^{1+c})$ as before.

Adding the space and time for the two phases together gives $O(n^{1+c})$ space and preprocessing time for graphs for which we can find a separator in $O(n^{1+c})$ time.

\subsection{Queries}\label{gnn:sec:queries}
Given a query node $q$, we find two sites: (a) the closest site to $q$ with paths restricted to the same side of the top-level partition as $q$, and (b) the closest site to $q$ with paths containing at least one separator node.
The paths considered between both cases cover all possible paths, so one of the two found sites is the overall closest site to $q$.
\begin{itemize}
	\item To find the site satisfying Condition (a), we can relay the query to the subgraph of the separator hierarchy containing $q$.
	This case does not arise if $q$ is a separator node.
	\item To find the site satisfying Condition (b), we need to find the shortest path from $q$ to any site, but only among paths containing separator nodes.
	Note that if the shortest path goes through a separator $s$, it should end at the site closest to $s$.
	Therefore, the length of the shortest path starting at $q$, going through $s$, and ending at any site, is $d(q,s)+d(s,\min(Q_s))$, where $\min(Q_s)$ denotes the element with the smallest key in $Q_s$.
	We can find the site satisfying Condition (b) by considering all the separator nodes and retaining the one minimizing this sum.
\end{itemize}

The time to find the site satisfying Condition (b) is $O(n^c)$, since there are $O(n^c)$ separator nodes to check and we do a find-minimum operation on a priority queue for each. We do not need to do any distance computation, as we precomputed all the needed distances.
Therefore, the time to find the two paths satisfying Conditions (a) and (b) can be analyzed by the recurrence
\[
T(n)\le T(2n/3)+O(n^c),
\]
where the $T(2n/3)$ term dominates the actual time for recursing in a single subgraph of the separator hierarchy.
The solution to this recurrence is $O(n^c)$,
so queries take $O(n^c)$ time.

We can implement a heuristic optimization for queries so that we do not need to check every separator node when searching for the site satisfying Condition (b).
During the preprocessing stage, we can sort, for each node $u$ in each graph $H$ of the separator hierarchy, all the separators in $H$ by distance from $u$.
This increases the space used by the data structure by a constant factor.
Then, during a query, after obtaining the site satisfying Condition (a), to find the site satisfying Condition (b), we consider the separator nodes in order by distance from the query node $q$.
Suppose that $p$ is the closest site found so far.
As soon as we reach a separator node $s$ such that $d(q,s)\geq d(q,p)$, we can stop and ignore the rest of separator nodes, since any site reached through them would be further from $q$ than $p$.
In our experiments (Section~\ref{gnn:sec:exps}), this optimization reduces the average query runtime by a factor between $1.5$ and $9.5$, depending on the number of sites.
It is more effective when there are many sites, as then the closest site is likely to be closer than many separators at the upper levels of the hierarchy.

\subsection{Updates}

Suppose that we wish to insert or delete a node $p$ to or from the set of sites $P$.
Note that, when we perform such an update, the structures computed during the site-independent preprocessing phase (the separator hierarchy and the computation of distances) do not change.
However, we need to add or remove $p$ (according to the type of update) to or from the priority queue $Q_s$ for every separator node $s$ in the top-level separator.
Moreover, if $p$ is not a separator node, we also need to update the priority queues for the subgraph containing $p$, recursively.

The time for an insertion is the same as for a query, since our priority queues support constant time insertions.
For deletions, the time to remove $p$ in all top-level priority queues is $O(\log k)$ time per priority queue, where $k$ is the number of sites, for a total time of $O(n^c\log k)$.
Again, if we formulate and solve a recurrence for the running time at all levels of the separator hierarchy, these times are dominated by the top-level term.

Next, we discuss how to improve the update time to $O(n^c\log\log n)$ with additional preprocessing.
For each separator node $s$, instead of using the distance from $s$ to $p$ as the key for a site $p$ in the priority queue $Q_s$, 
we can use the index of $p$ in the list of nodes sorted by distance from $s$.
That is, if the set of distances in sorted order from $s$ to the other nodes are
$d_1,d_2,d_3,\dots,$ with $d_1<d_2<d_3<\cdots$, we could replace these numbers by the numbers
$1,2,3,\dots$,
without changing the comparison between any two distances.
This replacement would allow us to use a faster integer priority queue in place of the priority queue. For instance, a van Emde Boas tree~\cite{vanEmdeBoas-SFCS-1975} maintains the minimum in a set of integer numbers between $1$ and $n$ in $O(\log\log n)$ time per insertion and deletion.
In order to use this optimization, we need to add the time to sort the distances in the preprocessing time, which increases to $O(n^{1+c}\log n)$ (assuming an $O(n\log n)$ time sorting algorithm is used).

We have completed the description and analysis of the data structure. Theorem~\ref{thm:graphnn} captures its runtime.

\begin{theorem}\label{thm:graphnn}
	Let $c$ be a constant with $0<c<1$, let $\mathcal{G}$ be a hereditary graph class with $O(n^c)$-size separators, and let $T(n)$ be the time needed to find a separator in an $n$-node graph from $\mathcal{G}$.
	Then, for any $n$-node graph from $\mathcal{G}$, there is a reactive nearest-neighbor data structure that uses $O(n^{1+c})$ space, with $\max{\left(O(n^{1+c}),T(n)\right)}$ preprocessing time, $O(n^c)$ query and insertion time, and $O(n^c\log k)$ deletion time, where $k$ is the number of sites.
	Alternatively, the data structure could have $\max{(O(n^{1+c}\log n), T(n))}$ preprocessing time, $O(n^c\log\log n)$ insertion and deletion time, and the same space and query time.
\end{theorem}

\section{Extensions and related data structures}

If we reformulate
and solve the recurrence equations for the case where
there is constant number of separator nodes ($c=0$), we obtain the space and runtimes shown in Table~\ref{tab:gnnsummary}.

The conga-line data structure~\cite{Eppstein2000} is a closest-pair data structure with $O(1)$ query time, $O(T(k)\log k)$ amortized insertion time, and $O(T(k)\log^2 k)$ amortized deletion time, where $T(k)$ is the time per operation (maximum between query and update) of a nearest-neighbor data structure maintaining $k$ sites. 
Another data structure~\cite{Eppstein1995} achieves the same runtimes, but for the bichromatic closest-pair problem.
Combined with our reactive nearest-neighbor data structure, we get the following result.

\begin{lemma}\label{lem:graphcp}
	Let $c$ be a constant with $0<c<1$, and let $\mathcal{G}$ be a hereditary graph class with $O(n^c)$-size separators.
	For any $n$-node graph from $\mathcal{G}$, there is a reactive closest-pair data structure and a reactive bichromatic closest-pair data structure with the space and runtimes shown in Table~\ref{tab:gdssummary}.
\end{lemma}

Finally, our reactive nearest neighbor data structure can be extended to directed graphs with the same asymptotic runtimes.
The only required change is to compute distances \emph{from} and \emph{to} every separator node.
To obtain the latter, we can compute the distances in the \textit{reverse graph}, i.e., the graph obtained by reversing the directions of all the edges.

\section{Experiments}\label{gnn:sec:exps}

In this section, we evaluate our data structure empirically on a real road network, the Delaware road network from the DIMACS data set~\cite{DIMACS}.
We consider the biggest connected component of the network, which has $48812$ nodes and $60027$ edges.
This data set has been planarized: overpasses and underpasses have been replaced by artificial intersection nodes.
Each trial in our experiment begins with a number of uniformly distributed random sites, and then performs 1000 operations.
We consider the cases of only queries, only updates, and a mixture of both (see Figure~\ref{gnn:fig:runtime}).
The updates alternate between insertions and deletions,
and the operations in the mixed case alternate between queries and updates.
We compare the performance of our data structure against a basic data structure that simply uses Dijkstra's algorithm for the queries.

\subsection{Implementation details}\label{gnn:sec:impl}

We implemented the algorithms in Java 8.\footnote{The source code with the implementation is available at \url{https://github.com/nmamano/NearestNeighborInGraphs}.}
We then executed them and timed them as run on an Intel(R) Core(TM) CPU i7-3537U 2.00GHz with 4GB of RAM, on Windows 10.

We implemented the optimization for queries described in Section~\ref{gnn:sec:queries}, and compared it with the unoptimized version in order to evaluate if its worth the extra space.
For updates, we used a normal binary heap, as these tend to perform better in practice than more sophisticated data structures.

A factor that affects the efficiency of the data structure is the size and balance of the separators.
Our hierarchy for the Delaware road network had a total of $504639$ nodes across $8960$ graphs up to $13$ levels deep.
Among these graphs, the biggest separator had $81$ nodes.
Rather than implementing a full planar separator algorithm to find the separators (recall that the data had been planarized), we choose the smallest of two simply-determined separators: the vertical and horizontal lines partitioning the nodes into two equal subsets.
While these are not guaranteed to have size $O(\sqrt{n})$, past experiments on the transversal complexity in road networks~\cite{Eppstein2009} indicate that straight-line traversals of road networks should provide separators with low complexity, making it unnecessary to incorporate a full planar graph separator algorithm.

When a separator partitions a graph in more than two connected components, we made one child per component.
Thus, our hierarchy is not necessarily a binary tree, and may be shallower.
We set the base case size to $20$.
At the base case, we perform Dijkstra's algorithm.
Experiments with different base-case sizes did not affect the performance significantly.

\subsection{Results}

\begin{figure}[t]
	\centering
	\includegraphics[width=0.999\linewidth]{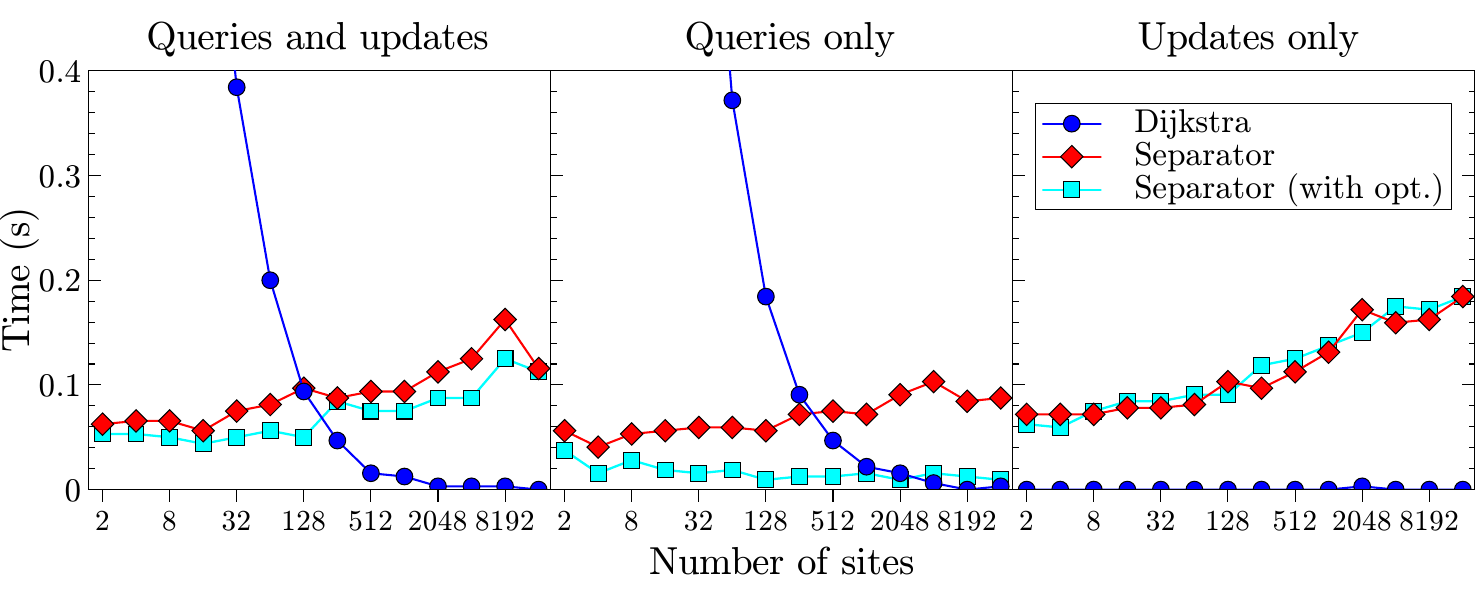} 
	\caption[Comparison of graph nearest neighbor data structures.]{Time needed by the data structures to complete 1000 operations in the Delaware road network~\cite{DIMACS} for a range of number of sites (in a logarithmic scale), excluding preprocessing time.
		Each data point is the average of 5 runs with different sets of random sites (the same sets for all the algorithms).}
	\label{gnn:fig:runtime}
\end{figure}

Figure~\ref{gnn:fig:runtime} depicts the results.
Table~\ref{gnn:tab:runtime} shows the corresponding data for the case of mixed operations, which is the case of interest in a reactive model.

\begin{itemize}
	\item The runtime of Dijkstra's algorithm is roughly inversely proportional to the number of sites, because with more sites it requires less exploration to find the closest one.
	Moreover, initialization and updates require virtually no time.
	Thus, this choice is superior for large numbers of sites, while being orders of magnitude slower when the number of sites is low (see Table~\ref{gnn:tab:runtime}).
	\item Our data structure based on a separator hierarchy is not affected as much by the number of sites.
	The update runtime only increases slightly with more sites because of the operations on larger heaps, as expected from its asymptotic runtime.
	The optimization, which reduces the number of separators needed to be checked, can be seen to have a significant effect on queries, especially as the number of sites increases: it is up to $9.5$ times faster on average with $2048$ sites.
	However, since it has no effect on updates, in the mixed model with the same number of updates and queries the improvement is less significant.
	\item The data structure requires a significant amount of time to construct the hierarchy.
	Our code constructed the hierarchy for the Delaware road network in around 15 seconds.
	Fortunately, this hierarchy only needs to be built once per road network.
	The limiting factor is the space requirement of approximately $O(n^{1+5})$, which caused us to run out of memory for other road networks from the DIMACS data set with over $10^5$ nodes.
\end{itemize}

\begin{table}[h!]
	\centering
	\begin{tabular}{l@{\hskip 0.3in}l@{\hskip 0.3in}l@{\hskip 0.3in}l}
		\# sites & Dijkstra            & Separator        & Separator (with opt.) \\ \hline
		2               & 3797 (3672 -- 3906) & 63 (47 -- 94)    & 53 (31 -- 94)        \\
		4               & 2303 (2203 -- 2359) & 66 (63 -- 78)    & 53 (47 -- 63)        \\
		8               & 1272 (1250 -- 1297) & 66 (47 -- 78)    & 50 (47 -- 63)        \\
		16              & 694 (641 -- 781)    & 56 (47 -- 63)    & 44 (31 -- 47)        \\
		32              & 384 (359 -- 406)    & 75 (63 -- 94)    & 50 (47 -- 63)        \\
		64              & 200 (172 -- 219)    & 81 (63 -- 94)    & 56 (47 -- 63)        \\
		128             & 94 (94 -- 94)       & 97 (94 -- 109)   & 50 (47 -- 63)        \\
		256             & 47 (47 -- 47)       & 88 (78 -- 94)    & 84 (78 -- 109)       \\
		512             & 16 (16 -- 16)       & 94 (94 -- 94)    & 75 (63 -- 78)        \\
		1024            & 13 (0 -- 16)        & 94 (94 -- 94)    & 75 (63 -- 78)        \\
		2048            & 3 (0 -- 16)         & 113 (94 -- 125)  & 88 (78 -- 94)        \\
		4096            & 3 (0 -- 16)         & 125 (109 -- 156) & 88 (78 -- 94)        \\
		8192            & 3 (0 -- 16)         & 163 (125 -- 188) & 125 (94 -- 156)      \\
		16384           & 0 (0 -- 0)          & 116 (109 -- 125) & 113 (94 -- 156)     
	\end{tabular}
	\caption[Comparison of graph nearest neighbor data structures.]{Time in milliseconds needed by the data structures to complete 1000 operations (mixed queries and updates) in the Delaware road network for a range of number of sites (in a logarithmic scale).
		Each data point is the average, minimum, and maximum, of 5 runs with different sets of random sites (the same sets for all the algorithms).}
	\label{gnn:tab:runtime}
\end{table}

\section{Conclusions}
We have studied reactive proximity problems in graphs, giving a family
of data structures for such problems. Tables~\ref{tab:gnnsummary} and~\ref{tab:gdssummary} summarize our theoretical results.
While we have focused on applications in geographic systems dealing with real-time data, the problems are primitive enough that they may arise in other domains of graph theory, such as network protocols.

We would like to explore other applications in the future.
New applications may require designing reactive proximity data structures for more general graph classes, i.e., classes without sublinear separators.
If finding exact nearest neighbors turns out to be too complex without sublinear separators, it would be interesting to design a reactive data structure supporting \textit{approximate} nearest-neighbor queries.

As discussed in Section~\ref{gnn:sec:impl}, an important factor in the runtime of any data structure based on separator hierarchies is the choice of separators.
It may be of interest to compare the benefits of a simpler but lower-quality separator construction algorithm versus a slower but higher-quality separator construction algorithm in future experiments.

\bibliographystyle{splncs03}
\bibliography{arxivbiblio}

\end{document}